# Reducing Technician Search Burden: A Multimodal RAG for Cessna 172 Maintenance Manual

Seongjun Ha[*]
*Purdue University, West Lafayette, Indiana, 47907, United States*

Md Rashedul Islam[†]
*Clemson University, Clemson, South Carolina, 29634, United States*

Gaurav Nanda[‡]
*Purdue University, West Lafayette, Indiana, 47907, United States*

and
Damon Lercel[§]
*Purdue University, West Lafayette, Indiana, 47907, United States*

**Proper use of the aircraft maintenance manual is essential for correct maintenance, providing procedures, diagrams, cautions, and specifications. However, technicians often avoid consulting it because it is difficult to navigate and time-consuming under strict schedules. Retrieval augmented generation (RAG) models have recently been introduced in aircraft maintenance, yet existing models focus solely on textual retrieval. This research therefore targeted the Cessna 172 Maintenance Manual (C172-MM), widely used in general aviation, and developed a multimodal manual retriever (MMR) capable of retrieving multimodal manual pages. Retrieval performance was evaluated using synthetic queries covering procedures, diagrams, caution/safety information, and specifications; the MMR achieved 93.37% recall@5. Beyond retrieval, a multimodal RAG (MRAG) pipeline was examined, in which retrieved pages were input to a vision-language model that generated responses to the synthetic queries, achieving 87.20% semantic similarity to ground-truth answers. Three practical feasibilities were also assessed: inference time, operational cost, and interpretability. Average retrieval time for five pages was 11.93 seconds and response generation took 4.95 seconds, at $0.0091 per query, while interpretability was validated through heatmap visualizations. These results indicate that the MRAG pipeline for the C172-MM can reduce the time technicians spend searching manuals and retrieving multimodal information.**

## Nomenclature

| | | |
|---|---|---|
| ATA | = | Air Transportation Association |
| C172-MM | = | Cessna 172 Maintenance Manual |
| C/S | = | Caution/Safety |
| DCG | = | Discounted Cumulative Gain |
| IDCG | = | Ideal Discounted Cumulative Gain |

[*]Graduate Research Assistant, School of Aviation and Transportation Technology.
[†]Graduate Research Assistant, Department of Industrial Engineering.
[‡]Assistant Professor, School of Engineering Technology.
[§]Associate Professor, School of Aviation and Transportation Technology.

| | | |
|---|---|---|
| IT | = | Input Tokens |
| LLM | = | Large Language Model |
| MMR | = | Multimodal Manual Retriever |
| MRAG | = | Multimodal Retrieval Augmented Generation |
| nDCG | = | Normalized Discounted Cumulative Gain |
| OT | = | Output Tokens |
| Q | = | Query |
| RAG | = | Retrieval Augmented Generation |
| s | = | Second |
| Spec | = | Specification |
| ViDoRe | = | Visual Document Retrieval Benchmark |
| VLM | = | Vision Language Model |

## I. Introduction

Aircraft technicians perform maintenance to ensure the continued airworthiness and safety of the aircraft. To perform maintenance correctly, technicians are required to consult and adhere to the procedures described in the manufacturer's maintenance manual or some other approved data, such as a service bulletin or airworthiness directive. For the remainder of this paper, the various types of approved data will be referred to as the 'maintenance manual' or 'manual' for brevity. The manual provides detailed instructions on the proper methods and standards for completing maintenance tasks, including removal, replacement, installation, adjustment, test, check, inspection, and cleaning [1]. Technicians can access the maintenance manual in two formats: a physical hard copy and a digital PDF version that can be viewed on a portable tablet or laptop [2]. The choice of which format to use often depends on each technician's personal preference and the specific task being performed.

Technicians use two main methods to locate the correct section of a maintenance manual for a specific task. When working with a physical hard copy of the manual, technicians typically rely on the Air Transportation Association (ATA) chapter classification system, which organizes content into sections formatted as *xx-yy-zz* along with a corresponding page range. In this structure, *xx* indicates the chapter covering a specific system, *yy* indicates a subsystem, and *zz* indicates a unit. Page numbers follow designated ranges: 1-99 are allocated for description and operation, while each major maintenance category, such as troubleshooting, maintenance practices, servicing, removal/installation, adjustment/test, inspection/check, cleaning/painting, and approved repairs – is assigned a separate 100 page block, respectively [3,4]. By following this standardized structure, technicians can quickly navigate to the appropriate section of the manual and the desired maintenance task.

When using the digital PDF version, technicians often depend on the search function, entering keywords to quickly locate maintenance tasks. However, a keyword search is not always effective across different aircraft manufacturers because terminology for the same component may vary. For example, one manufacturer may refer to a part as a "*washer*" while another may call it a "*spacer*." Additionally, searching for overly broad terms, such as "*bolt*," can also produce an overwhelming number of results. According to Sadri [5], about 2.4 million hardware parts are needed to build a Boeing 747 aircraft. Keyword search results may still require substantial scrolling and manual filtering before the relevant passage is found due to the large volume of content in the maintenance manual.

Difficulty locating information in the manual often turns the process into a time-consuming and frustrating task, which can lead some technicians to begin work without fully reading the maintenance manual beforehand. This may contribute to incorrect or incomplete maintenance actions. For example, Hobbs [6] pointed out that aircraft technicians spend 20 to 40 percent of their time searching for information in the maintenance manual. Avers et al. [7] found that technical data for maintenance tasks are often not located in a single place; instead, technicians are required to consult multiple manuals and multiple sections within those manuals. For example, Avers et al. [7] noted that performing maintenance on the A320 anti-skid transducer connector required 62 maintenance manual sections and another 11 additional supplement documents; the combined weight of these documents was approximately 6.25 pounds.

Technicians often need accurate information immediately; however, when retrieving that information becomes time-consuming and the effort outweighs the perceived benefit [8,9], technicians may be less motivated to search the maintenance manuals. From the researchers' perspective, when information is easily retrieved, reluctance to search through maintenance manuals is significantly reduced. These challenges highlight the need for more effective information retrieval tools that support technicians during maintenance tasks [10].

To address these limitations, this study developed a multimodal manual retriever (MMR) and a multimodal retrieval augmented generation (MRAG) pipeline tailored to the Cessna 172 Maintenance Manual (C172-MM). The MMR retrieved relevant maintenance pages, and the MRAG pipeline generated a response to the technician's query based on those retrieved pages. By extending prior aviation maintenance RAG models that relied primarily on text-only retrieval, the proposed MRAG pipeline better supported the multimodal information needs of aircraft maintenance and reduced the manual search burden.

## II. Literature Review

### A. Factors Behind Technicians Not Using Maintenance Manual

The first step in performing any maintenance task is to obtain a work order from a supervisor and then review a maintenance manual to properly conduct the required maintenance procedures [2]. The maintenance manual is intended to provide the necessary instructions needed to perform required maintenance tasks, including removal, replacement, installation, adjustment, test, check, inspection, and cleaning presented through procedural steps, caution or safety notes, diagrams, and specifications [1]. The manual typically does not specify common tools needed to complete a task, such as the particular type of screwdriver, a flashlight, or wrench and socket sizes, so technicians are required to draw on their own experience to fill these gaps. In practice, for example, before beginning a task, technicians are required to consult the disassembly section to identify which components need to be removed and review the cleaning section when applicable. After completing the assembly, technicians then refer to the inspection section to determine which conditions to look for, what constitutes an abnormal or defective state, and how to evaluate component serviceability based on guidance provided in the maintenance manual [2].

However, several studies have shown that technicians sometimes do not fully read the maintenance manual when performing a task [7–9,11–13]. This behavior has been attributed to multiple factors. One commonly reported reason is that technicians often perceive their personal experience, habitual practice, or preferred techniques as more effective than procedures prescribed in the maintenance manual [9,14–16]. In addition, technicians report that maintenance manuals often have low usability, such as poor readability, unclear content structure, disorganized layout, and difficulties with physical handling, which makes it challenging and time consuming to locate the correct information [9,14,17,18]. The Federal Aviation Administration [19] also noted that technicians frequently face schedule pressure, which may further discourage thorough manual reading. Zafiharimalala et al. [9] found that strictly adhering to the maintenance manual procedures is not practical for technicians. Following all steps as written would make it highly improbable that a technician would complete the maintenance task within the scheduled time. In addition, Avers et al. [7] noted that maintenance manual information is distributed across multiple sections of the manual and that technicians spend 20 to 40% of their time consulting maintenance information.

## B. Efforts in Improvement of Aircraft Maintenance Manual

While aircraft maintenance manuals were originally developed as physical, paper-based documents and are still used in that form today, the study and design of maintenance manuals have continued to evolve. Early research identified several limitations of paper manuals, including poor readability, handling difficulties, and challenges in keeping revision up to date [18]. As demonstrated in Patel et al. [20], the usability of maintenance documentation was enhanced by redesigning work cards with attention to information readability, contextual relevance, organizational structure, and physical handling factors. Additionally, the introduction of simplified English provided significant benefits for non-native English speaking technicians, helping them more easily understand procedural instructions [21].

Drury et al. [22] also found that the introduction of portable computers improved manual usability by enabling direct links to related documents. Gunes et al. [12] developed an interactive and intuitive maintenance manual designed for use on portable devices. Additionally, Liang et al. [23] developed an online maintenance assistant platform designed to reduce human factor errors during tasks and improve technicians' overall performance. Jo et al. [24] applied an ontology-based approach to reduce ambiguity in procedural interpretation, which helped decrease the time required for technicians to prepare maintenance tasks.

More recently, studies by Hou et al. [24], Signé et al. [25], and Jo [27] have explored the development of aircraft maintenance assistants using retrieval-augmented generation (RAG). RAG is motivated by limitations of relying solely on an LLM for domain-specific correctness, including the need for frequent updates and the risk of hallucinated responses [28,29]. RAG addresses these limitations through three components: (a) retrieval of relevant documents from a vector database, (b) an LLM generator that produces a response, and (c) augmentation, which determines how and when the retrieved information is integrated into the generation process [28]. For example, Jo [27] reported a 98% reduction in technicians' manual search time.

However, the three existing RAG models proposed by Hou et al. [25], Signé et al. [26], and Jo [27] are constrained to text-based retrieval, which may reduce their effectiveness, as technicians often rely on visual information such as diagrams, illustrations and tables during maintenance. Therefore, an MRAG approach that retrieves text, diagrams, and tables can provide greater support for technicians by offering the visual context needed to reduce uncertainty in textual explanations, improve understanding of complex procedures, and minimize the risk of misinterpretation [12,18,22,24].

## III. Research Objective

Building on the literature reviewed above, the goal of this research is to enable technicians to obtain multimodal maintenance manual pages and direct response, which potentially reduces the manual search burden during maintenance tasks. Therefore, this study aims to examine the feasibility and efficacy of MMRs by developing an MMR tailored to the C172-MM and evaluating its performance in retrieving maintenance manual pages. Based on the retrieval performance of the MMR, it is then incorporated into an MRAG pipeline, which generates responses to technician's query. The generated responses are compared with ground truth answers. Then, three aspects of the MRAG pipeline's practical feasibility, retrieval and response generation inference time, operational cost, and MMR interpretability are examined.

## IV. Methodology

Fig. 1 provides an overview of the study methodology. Following the research objective described above, the methodology is organized into five components, which focused on C172-MM: (1) development of the MMR, (2) development of the MRAG pipeline, (3) creation of synthetic test dataset, (4) assessment of MMR, and (5) assessment of MRAG pipeline and practical feasibilities such as retrieval and response generation time, operational cost, and MMR interpretability.

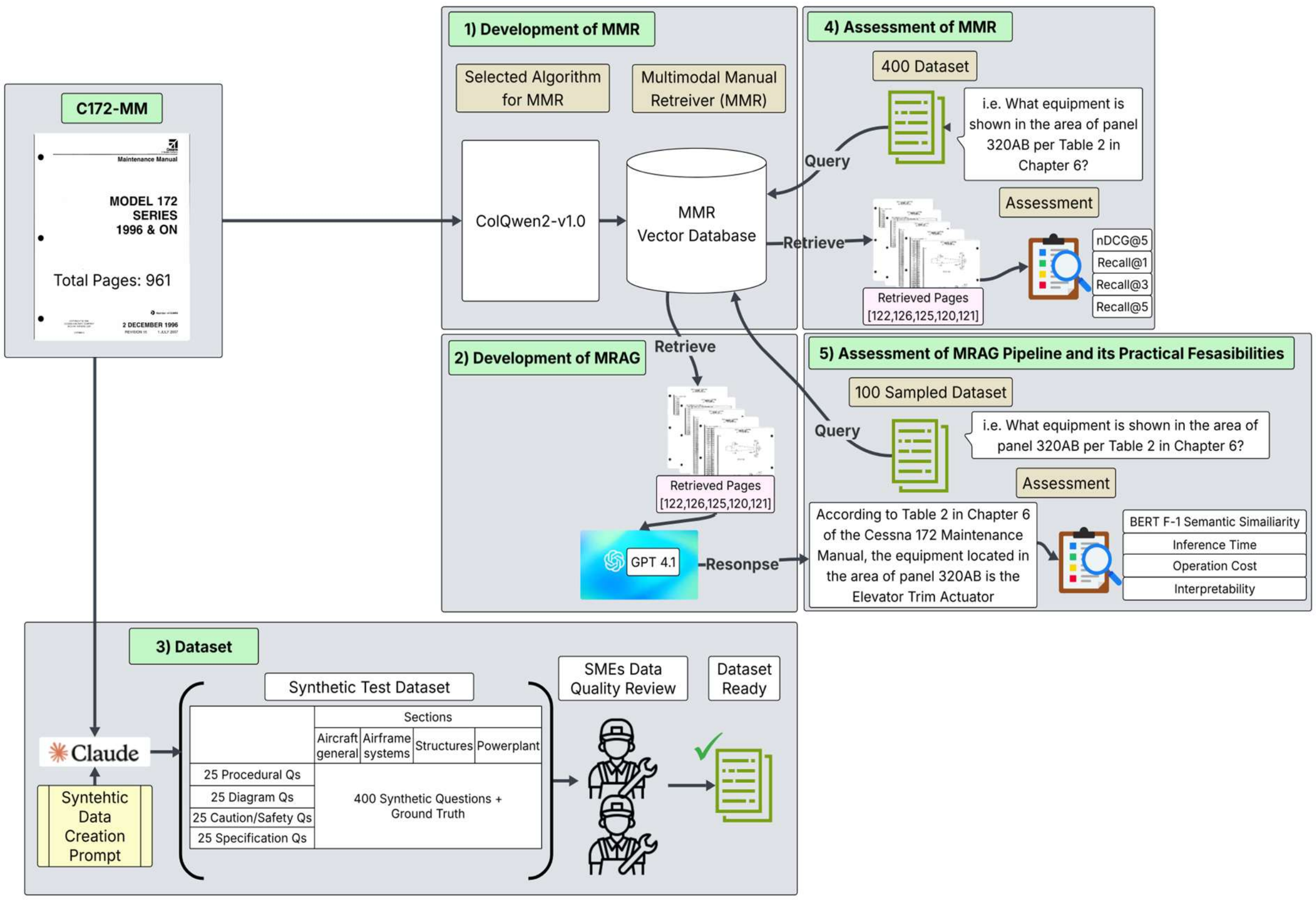

**Fig. 1: Methodology Overview**

## A. Development of Multimodal Manual Retriever (MMR)

Traditional RAG model development requires document pre-processing, including content extraction and parsing [28]. To perform this preprocessing efficiently, deep learning models, such as optical character recognition, are commonly used to detect and extract textual content from documents. A comprehensive toolkit like Docling [30] further streamlines this process by providing integrated support for content extraction, document conversion, and parsing across multiple document formats into a unified and structured representation. The extracted content is then converted into embeddings, numerical vector representations that allow semantically relevant content to be retrieved for a given query [28]. In a typical MRAG pipeline, each data type, such as text, images and tables, is embedded separately. However, the MRAG performance is not guaranteed because the image and text embeddings are produced separately, which often results in weak cross modal alignment [31].

Recognizing these limitations, ColPali model [32] introduced simplified multimodal retrieval architecture. The ColPali model embeds each document page as an image, avoiding the multi-stage preprocessing typical of textual only retrieval pipelines while maintaining promising retrieval performance. A key contribution of the ColPali model can be explained in two components. The "*Col*" component represents the adaptation of the model's multimodal representations specifically for retrieval tasks, leveraging ColBERT style late interaction mechanism. The "*Pali*" component refers to using PaliGemma as the vision language model (VLM) embedding backbone, which has already internalized robust text-in-image understanding through extensive pretraining. The ColPali model represents each document page as an image and computes fine-grained similarity scores by applying MaxSim to match each query token embedding against all document patch embedding. This enables direct token-to-patch alignment at retrieval time.

As more advanced VLM backbones are developed and available, the ViDoRe benchmark [32,33] evaluated multimodal retrieval performance across different VLM backbones and multiple domain specific datasets. This study developed an MMR for C172-MM using ColQwen2-v1.0 architecture, which employs Qwen 2-VL-2b-instruct as its VLM backbone. This model has a relatively smaller parameter size compared with other high-performing models and still achieves high retrieval accuracy based on the ViDoRe benchmark. Given the limited computational resources for this research, ColQwen2-v1.0 offered a more practical balance between accuracy, inference time, and computational efficiency. To develop the MMR for the C172-MM, the researchers used Google Colab with an H100 GPU to embed the entire C172-MM PDF.

### B. Development of Multimodal Retrieval Augmented Generation (MRAG) Pipeline

While the ColPali model [32] primarily focused on developing an architecture for multimodal retrieval, the MMR above does not have the ability to directly respond to a technician's query. Instead, it returns only the retrieved C172-MM pages without explanation, meaning that technicians still need to manually read and interpret the retrieved responses. Nevertheless, the ColPali model highlighted the promising capabilities of VLMs for page as image query answering, as demonstrated by Ma et al. [34]. Therefore, the researchers constructed an MRAG pipeline for the C172-MM in which the MMR's top-$k$ retrieved pages are input into a VLM. In this study, $k$ was set to five to match the retrieval evaluation setting and to maximize the likelihood that the correct evidence page is included. GPT 4.1 was selected as the VLM.

### C. Dataset Creation

To evaluate the MMR and the MRAG pipeline, both tailored to the C172-MM, this study constructed a synthetic test dataset based on the 1996 and later series of the maintenance manual [3], consisting of 961 pages in PDF format. In developing a synthetic test dataset, queries and ground truth answers are required, and two perspectives are considered to ensure comprehensive and diverse assessment for the C172-MM: 1), the organizational structure of the maintenance manual, reflected by its major chapter groupings, and 2) the task-oriented information needs of aircraft technicians, as reflected in the types of information commonly needed during maintenance tasks.

For the organizational structure of the maintenance manual, the C172-MM is organized into four major chapters, such as: aircraft general, airframe systems, structures, and powerplant, as shown in Table 1.

**Table 1: Major Chapters of C172-MM**

| Major Chapters | Aircraft General | Airframe Systems | Structures | Powerplant |
|---|---|---|---|---|
| Chapter–Title | 4 – Airworthiness | 20 - Standard practices - airframe | 51 – Structures | 61 – Propeller |
| | 5 - Time Limits/Maintenance checks | 21 - Air conditioning | 52 - Doors | 70 – Standard practices – Engine |
| | 6 - Dimensions and areas | 22 - Auto flight | 55 - Stabilizers | 71 – Powerplant |
| | 7 - Lifting and shoring | 23 - Communications | 56 - Windows | 73 – Engine fuel and control |
| | 8 - Leveling and weighing | 24 - Electrical power | 57 - Wings | 74 – Ignition |
| | 9 - Towing and taxiing | 25 - Equipment/furnishings | | 77 – Engine Indicating |
| | 10 - Parking, mooring | 27 - Flight controls | | 78 – Exhaust |
| | 11 -Placards and markings | 28 - Fuel | | 79 – Oil |
| | 12 -Servicing | 32 - Landing gear | | 80 – Starting |
| | | 33 - Lights | | |
| | | 34 - Navigation | | |
| | | 37 - Vacuum | | |
| | | 39 – Electric/Electronic panels and multi-purpose parts | | |
| | | 91 – Charts and wiring diagrams | | |

Prior work on aviation maintenance documentation has emphasized the need to distinguish between different categories of information such as directive information, reference information, warnings and cautions, and procedures to improve usability [20]. Building on this perspective, the researchers reviewed the structure and content of the C172-MM and identified four categories of task-oriented information needs: 1. *Procedural information*: specific maintenance procedures; 2. *Diagram information:* figures, diagrams, tables or illustrations; 3. *Caution/Safety* *information*: warnings, cautions, safety, or notes during maintenance tasks; 4. *Specification information*: technical specifications, limits, or configuration details.

Based on these two perspectives, a prompt template was developed to guide Claude Sonnet 4.6 [35] in generating the synthetic test dataset, as shown in Fig. 2. Using this template, the Claude Sonnet 4.6 created 100 queries for each major chapter group, with 25 queries allocated to four task-oriented information, resulting in a total of 400 queries in the synthetic test dataset. In addition, ground truth answers were derived directly from the maintenance manual. For each query, the dataset also records the corresponding chapter reference, manual page number, and PDF page location to ensure traceability to the source document.

```
You will read the Cessna 172R Maintenance Manual.
For each of the following chapter ranges:
- Chapter 5–12
- Chapter 20–37
- Chapter 51–57
- Chapter 61–80
Create 100 queries per chapter range, divided into:
- 25 procedural queries
- 25 diagram queries
- 25 caution/safety queries
- 25 specification queries
For each query, include:
- "query"
- "answer" (ground truth from the manual)
- "chapter_reference, page number"
- “PDF page location”

Output everything in valid JSON.
```

**Fig. 2: Prompt Template for C172-MM Synthetic Test Dataset Creation**

Since the synthetic test dataset is generated dataset created using the Claude Sonnet 4.6, the researchers conducted quality checks with two subject matter experts in aircraft maintenance research and two FAA certified airframe and powerplant technicians with Cessna 172 maintenance experience. These subject matter experts reviewed the generated queries and their corresponding ground truth answers to assess whether Claude 4.6 accurately reflected the types of queries that technicians commonly encounter during real maintenance tasks, whether the dataset adhered to the instructions specified in the prompt template, and whether the ground truth information was correctly produced. This validation step ensured that the synthetic test dataset was reliable, structurally diverse, and aligned with authentic maintenance practices. If any of the generated queries were found to be unrealistic or not representative of actual maintenance scenarios, the subject-matter experts revised them accordingly. As shown in Fig. 3, the distribution of the 400 synthetic queries of the C172-MM chapter is presented. This synthetic test dataset was then used in the following two assessments.

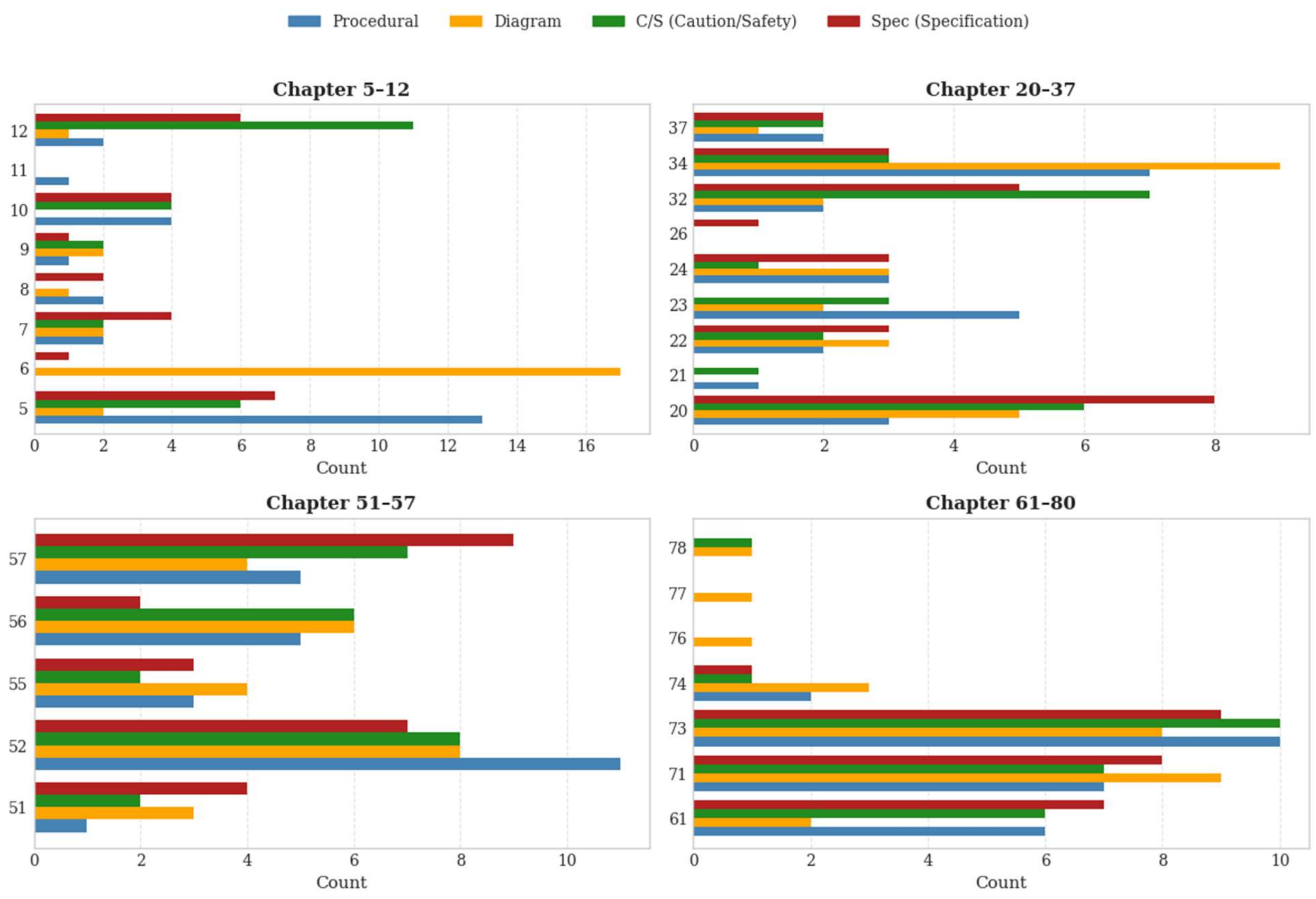

**Fig. 3: Dataset Distributions by Major Chapter in the C172-MM**

## D. Assessment of Multimodal Manual Retriever (MMR)

The first assessment evaluated the MMR for C172-MM retrieval performance using two metrics: normalized discounted cumulative gain (nDCG@*k*) and recall@*k* on a dataset of 400 queries with corresponding ground truth PDF page.

The researchers assessed ranking quality using nDCG@*k* [36], a commonly adopted evaluation metric in information retrieval that measures how closely retrieval MMR's ranked pages align with a correct ground truth ordering. nDCG normalizes the discounted cumulative gain (DCG) by the ideal DCG (IDCG), producing a score between 0 and 1 that reflects the relative quality of the ranking. DCG applies a logarithmic discount to lower ranked results, as noted in Equations (1) to (3), ensuring that highly relevant items appearing earlier in the list contribute more to the overall score than those placed lower. Top-*k* for nDCG@k was set to five. Additionally, recall@*k* [37] was used, which measures the number of correctly retrieved pages in the top-$k$ divided by the total number of relevant pages in the ground truth. This metric evaluated how completely the MMR retrieves the relevant pages, as defined in Equation (4). Additionally, by varying the value of $k$ = 1, 3 and 5, the researchers examined how likely the correct page appears within the top-$k$ results.

$$DCG@k = \sum_{i=1}^{k} \frac{2^{rel}\sigma(i) - 1}{\log_2(i+1)} \tag{1}$$

$$IDCG@k = \sum_{i=1}^{k} \frac{2^{rel}i - 1}{\log_2(i+1)} \tag{2}$$

$$nDCG@k = \frac{DCG@k}{IDCG@k} \tag{3}$$

$$Recall@k = \frac{Number\ of\ correctly\ retrieved\ outputs\ in\ top\ k}{Ground\ truth\ of\ ks} \tag{4}$$

**E. Assessment of Multimodal RAG (MRAG) Pipeline and its Practical Feasibilities**

This second stage assessment aims to assess the MRAG pipeline for C172-MM performance and its practical feasibilities. Four perspectives are assessed: 1. BERT-based F-1 semantic similarity between MRAG responses and ground truth answers; and three additional practical feasibilities are assessed: 2. MMR's retrieval and MRAG's response generation inference time, 3. MRAG pipeline operational cost, 4. MMR interpretability.

From the dataset of 400 queries and corresponding ground truths, the researchers sampled 25 queries from each task oriented query type, resulting in a total of 100 queries ($N$=100).

For the first assessment, the responses generated by the VLM were evaluated against the ground truth textual answers using the BERTscore metric for semantic similarity [38], which computes semantic similarity based on token embeddings extracted from RoBERTa [39]. To define the components of equations, let $X = \{x_i\}$ denote the set of token embeddings derived from the ground truth sequence, and let $\hat{X} = \{\hat{x_j}\}$ denote the set of token embeddings obtained from the VLM's response tokens. Here, $i$ and $j$ index the ground truth and predicted tokens, respectively. The semantic similarity between two tokens is quantified by the cosine similarity of their contextual embeddings. The BERT-based recall measures how well the semantic content of the ground truth is preserved in the predicted response tokens. For each ground truth token embedding $x_i$, the most similar predicted token embedding $\hat{x_j}$ is identified, and the recall is computed as in Equation (5). Conversely, the BERT-based precision measures how much of the response tokens are semantically supported by the ground truth. For each predicted token embedding $\hat{x_j}$, the most similar ground truth token embedding $x_i$, is selected, and the precision is computed as in Equation (6). The BERT-based F-1 score is computed as the harmonic means of BERT recall and BERT precision, as shown in Equation (7). A higher F-1 value indicates high semantic alignment between the ground truth and the VLM's response tokens.

$$Recall_{BERT} = \frac{1}{|x|} \sum_{x_i \in X} \max_{\widehat{x_j} \in \hat{X}} x_i^{\top} \widehat{x_j} \tag{5}$$

$$Precision_{BERT} = \frac{1}{|\hat{x}|} \sum_{\widehat{x_j} \in \hat{X}} \max_{x_i \in X} x_i^{\top} \widehat{x_j} \tag{6}$$

$$F-1_{BERT} = 2 \times \frac{Precision_{BERT} \times Recall_{BERT}}{Precision_{BERT} + Recall_{BERT}} \tag{7}$$

Beyond the semantic similarity assessment of the MRAG pipeline's responses described above, three practical feasibilities were evaluated.

1. MMR's Retrieval and MRAG's Response Generation Inference Time: The researchers measured MMR's C172-MM retrieval time and MRAG's C172-MM response generation time using CPU only execution to simulate deployment in computationally limited environments. This experiment was conducted in the Google Colab using an Intel® Xeon CPU @ 2.20 GHz with 2 vCPUs (1 core, 2 threads per core).
2. MRAG Operational Cost: The retrieved maintenance manual pages from the MMR were input into GPT 4.1 as the VLM, and the cost of generating the MRAG response was analyzed. GPT 4.1 charges usage in two components: the cost of input tokens, which are incurred when multimodal data is input into the GPT 4.1, and the cost of output tokens, which are incurred when the GPT 4.1 generates a response. To more comprehensively measure operational cost, researchers computed the average number of input and output tokens consumed per query and the total input and output tokens usage across the dataset. Based on the token usage, the operational cost was computed according to GPT 4.1's pricing, which is $ 0.000002 per input token and $ 0.000008 per output token [40].
3. MMR Interpretability: The MMR supports interpretability through heatmaps that visualize how individual query tokens align with the text in the retrieved maintenance manual pages. By highlighting which parts of the input contribute to the MMR's retrieval decisions, these heatmaps provide transparency and explainability.

## V. Results

### A. Performance of MMR for C172-MM

Using the MMR for C172-MM, the nDCG@5 results showed that the four chapter groups performed consistently well as shown in Table 2: aircraft general (chapters 5 to 12) achieved 83.62%, aircraft systems (chapters 20 to 37) achieved 84.01%, structures (chapters 51 to 57) achieved 87.26%, and powerplant (chapters 61-80) achieved 78.66%, resulting in an overall performance of 83.38% across the entire manual. This indicates that the MMR successfully placed a highly relevant maintenance manual page within the top five ranks for the given queries.

**Table 2: nDCG and Recall Performance Across Chapter Groups, Entire Manul, and Query Types**

| Metrics | Chapters | | | | Entire Manual | Query Types | | | |
|---|---|---|---|---|---|---|---|---|---|
| | 5-12 | 20-37 | 51-57 | 61-80 | | Procedural | Diagram | C/S[a] | Spec[b] |
| nDCG@5 | 0.8362 | 0.8401 | 0.8726 | 0.7866 | 0.8338 | 0.8629 | 0.6963 | 0.8723 | 0.9039 |
| Recall@1 | 0.6900 | 0.7200 | 0.7400 | 0.6083 | 0.6895 | 0.7350 | 0.4733 | 0.7750 | 0.7750 |
| Recall@3 | 0.9050 | 0.8800 | 0.9150 | 0.8250 | 0.8812 | 0.9150 | 0.7550 | 0.9050 | 0.9500 |
| Recall@5 | 0.9250 | 0.9250 | 0.9650 | 0.9200 | 0.9337 | 0.9500 | 0.8400 | 0.9450 | 1.0 |

[a] Caution/Safety [b] Specification

Using the recall metric, the researchers varied the top-$k$ setting ($k$ = 1, 3, 5) to evaluate how likely MMR was to return the correct maintenance manual page within the top-$k$ retrieved results for a given query. Across the entire manual, the MMR achieved 68.95%, probability that the first retrieved maintenance manual page matched the ground truth page for the synthetic test dataset queries. As the top-$k$ value increases, the MMR retrieved more candidate maintenance manual pages, which improved the likelihood of including the correct pages. With $k$ = 3, the MMR correctly retrieved the relevant maintenance manual pages for 88.12% of ground truth pages. With $k$ = 5, the MMR retrieved the correct pages for 93.37% of ground truth pages. This represents a 24.42 % point increase in recall from $k$ = 1 to $k$ = 5, indicating a substantial improvement in the likelihood of retrieving the ground truth pages as more candidates are considered.

More specifically, based on the overall performance across the entire manual, the researchers examined how the results varied by each of the following query types: procedural, diagram, caution/safety, and specification. As shown in Table 2, specification queries performed well in nDCG@5 ranking accuracy within the top five retrieved outputs, achieving a score of 90.39%. At recall@1, the MMR achieved 73.50% recall for procedural queries, 47.33% for diagram queries, 77.50% for caution/safety queries, and 77.50% for specification queries. When the recall top-$k$ setting increased to $k$ = 5, the MMR achieved 95.00% recall for procedural queries, 84.00% for diagram related queries, 94.50% for caution/safety queries, and 100% for specification queries, indicating recall improvements of 8.71%. 14.37%, 7.27% and 9.61%, respectively. Notably, diagram related queries showed the largest improvement, and all query types showed substantial improvements in recall as the top-$k$ value increased.

**B. Performance of MRAG Pipeline for C172-MM and Its Practical Feasibilities**

Based on the strong maintenance manual page retrieval performance of 93.37% at recall@5, the researchers adopted $k = 5$ as the top-$k$ setting. The MRAG pipeline for C172-MM using the GPT 4.1 VLM was then assessed with this setting, and the results are presented in Table 3.

**Table 3: MRAG for C172-MM Performance with $k = 5$ Setting and its Practical Feasibilities**

| Metrics | Query Types ($n$=25) | | | | Overall |
|---|---|---|---|---|---|
| | Procedural | Diagram | C/S | Spec | ($N$=100) |
| Average BERT F-1 semantic similarity per Q[c] | 0.8656 | 0.8655 | 0.8942 | 0.8629 | 0.8720 |
| Practical Feasibilities | | | | | |
| Average MMR retrieval time (s[b]) | 11.6167 | 13.9279 | 10.6481 | 11.5185 | 11.9278 |
| Average MRAG response time (s[b]) | 6.1623 | 4.2313 | 4.7574 | 4.6579 | 4.9522 |
| Average MRAG IT[a] per Q[c] | 3889 | 3894 | 3892 | 3899 | 3893.5 |
| Average MRAG OT[d] per Q[c] | 327 | 94 | 125 | 105 | 162.75 |
| Average MRAG Cost[e] per Q[c] ($) | 0.0103 | 0.0085 | 0.0087 | 0.0086 | 0.0091 |
| Total MRAG IT[a] | 97225 | 97350 | 97300 | 97475 | 389350 |
| Total MRAG OT[d] | 8175 | 2350 | 3125 | 2625 | 16275 |
| Total MRAG Cost[e] ($) | 0.2599 | 0.2135 | 0.2196 | 0.2160 | 0.9090 |

[a] input tokens [b] seconds [c] query [d] output tokens [e] Based on OpenAI GPT-4.1 pricing: $0.000002 per IT, $0.000008 per OT.

The average BERT-based F-1 semantic similarity was 87.20%. This indicated that the MRAG responses are highly semantically aligned with the ground truth answers from the C172-MM. Across query types, the MRAG responses' BERT-based F-1 semantic similarity was relatively consistent for the C172-MM.

In terms of MMR retrieval time and MRAG response time, the overall average MMR response time was 11.9278 seconds to retrieve five maintenance manual pages from the MMR and 4.9522 seconds to generate a response for each query. In terms of MRAG operational cost, each query consumed 3893.5 input tokens and 162.75 output tokens, resulting in a per query cost of approximately $ 0.0091. For the entire test dataset ($N$ = 100), the MRAG consumed 389350 input tokens and 16275 output tokens, with a total cost of $ 0.9089.

The researchers then analyzed the MMR retrieval time, the MRAG pipeline's response time, and its operational cost. The MMR spent noticeably longer on diagram queries, averaging 13.93 seconds, which is about two seconds above the overall average. This study assumes that diagram retrieval is more complex because diagrams contain limited textual information. Typically, a diagram includes only figure names, legends, or minimal descriptive text, making it more challenging for the MMR to extract meaningful semantic cues compared to other query types. Additionally, for procedural queries, the MRAG spent an average of 6.1623 seconds per query to generate a response, approximately one second longer than average, which also resulted in a higher GPT token cost of $ 0.0103. This is

because procedural queries require longer, more detailed responses, which increase output token usage and result in the highest MRAG operation cost and response generation inference time. Since GPT 4.1 output tokens are four times more expensive than input tokens, longer responses directly raise generation cost. In contrast, diagram queries typically require shorter responses, resulting in lower output token consumption. Table 4 presents example results for each query type.

**Table 4: Example Results for Each Query Type**

| | (A) Procedural | (B) Diagram | (C) C/S | (D) Spec |
|---|---|---|---|---|
| Query | Describe the procedure for towing the Cessna 172 when no tow bar is available. | Which figure shows the weatherstrip installation details including the 0.25-inch notch location at the door low point? | What NOTE applies to individual two-piece injection nozzles during removal for cleaning? | What is the torque range for a 1/2-inch tubing O.D. with steel tubing and steel inserts? |
| Ground truth | Press down on the tailcone at a bulkhead to raise the nose wheel off the ground. With the nose wheel clear, pivot the airplane about the main wheels to turn it. | Figure 201, Sheet 3 – 'Cabin Door Installation', Section 52-10-00, Page 204. Shows View A-A and View B-B with weatherstrip perimeter application and 0.25-inch notch at door low point. | NOTE: Individual two-piece nozzles should be kept as MATCHED ASSEMBLIES (do not mix components from different nozzles). | 450 to 500 inch-pounds |
| Ground truth M/M page | 134 | 788 | 884 | 246 |
| MMR Retrieved Pages (*k*=5) | [134, 133, 132, 142, 135] | [788, 785, 830, 814, 831] | [884, 881, 584, 286, 875] | [246, 245, 244, 498, 948] |
| MRAG's Response | When towing a Cessna 172 without a tow bar, follow this procedure as described in the maintenance manual:<br>1. **Use Wing Struts and Landing Gear Struts:** Move the airplane by hand using the wing struts and landing gear struts as push points.<br>2. **Raising the Nose Wheel:** To maneuver the airplane without a tow bar, press down on the tailcone at a bulkhead to raise the nose wheel off the ground.<br>**[shortened…]** | The weatherstrip installation details, including the 0.25-inch notch location at the door low point, are shown in **Figure 201, View B-B** on the first image. This figure specifically illustrates the placement of the weatherstrip and the 0.25-inch notch at the bottom of the door structure. | The NOTE that applies to individual two-piece injection nozzles during removal for cleaning is:"Individual two-piece nozzles should be kept as matched assemblies."(Source: Image 1, Section 8.B.(1)) | For a 1⁄2-inch tubing O.D. with steel tubing and steel inserts, the torque range is **450 to 500 inch-pounds**. Reference: Table 205, under the column "Flared or Flareless fitting with Steel Tubing, and Hose with Steel Inserts," for 1/2" tubing O.D. |
| F-1 BERT | 0.8783 | 0.8986 | 0.8698 | 0.8661 |
| MMR retrieve time | 7.584 | 12.0492 | 9.403 | 12.3832 |
| MRAG response generation time | 8.5048 | 4.1857 | 4.9776 | 3.4577 |
| Operation Cost | 0.009576 | 0.008306 | 0.008156 | 0.008362 |

Based on the example results for each query type, the researchers conducted an MMR interpretability analysis. Fig. 4 presents heatmaps for the four query types: (A) Procedural, (B) Diagram, (C) Caution/Safety, and (D) Specification. These heatmap visualizations show how the query tokens align with the corresponding textual regions in the retrieved maintenance manual pages, illustrating the MMR's attention and supporting the interpretability of its retrieved pages.

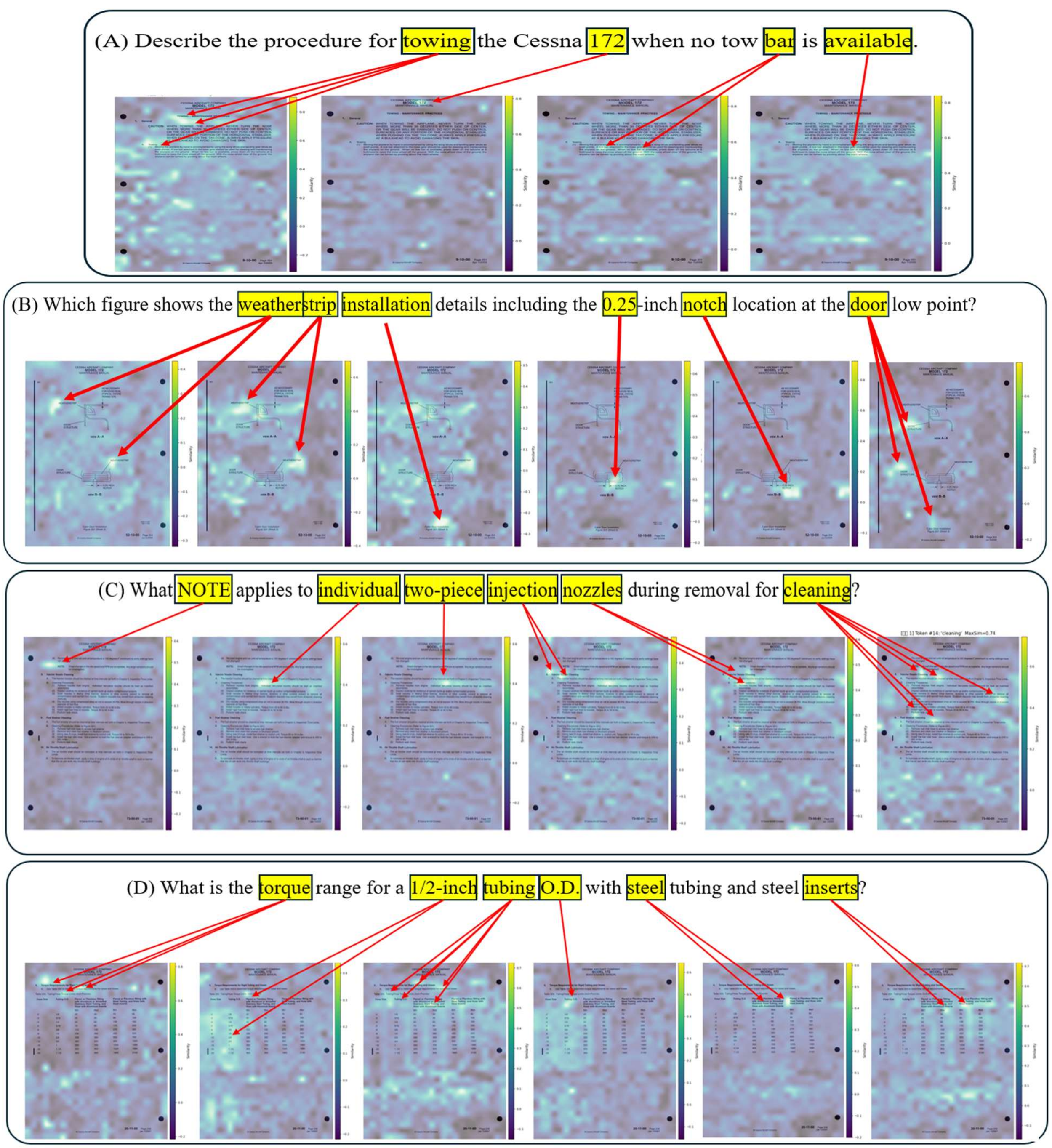

*Note*. Although each query contains more tokens, the heatmaps in Fig. 4 display only the tokens that are aligned with the retrieved pages, making the MMR's input-output connections clear.

**Fig. 4: Heatmap Visualizations for Query Tokens Corresponding to Retrieved Pages. (A), (B), (C), and (D) show examples. *Note.* The zoomed-in version can be found in Appendix A.**

## VI. Discussion

### A. Cascading Error Caused by MMR Failure

Although the MMR achieved 92.87% at recall@5, it still retrieved irrelevant maintenance manual pages. These irrelevant pages propagated into the response of the MRAG pipeline, causing VLM to generate an inaccurate response. This case represents a cascading error effect. As shown in Table 5, the MMR retrieved five C172-MM pages for the query, none of which matched the ground truth page. These incorrect retrieval pages were then input into the VLM, leading the final answer to deviate from the correct ground truth response.

For example, while page 337 contained procedures for VHF antenna removal and installation, the VLM hallucinated procedures based on page 338, which only included a diagram of the VHF communication antenna installation. The resulting MRAG response is shown in Table 5.

**Table 5: Cascading Error: MMR Retrieved Pages Propagate to Incorrect MRAG Response**

| Query | MMR Retrieved Pages | MRAG Response |
|---|---|---|
| "How is a VHF antenna removed from the Cessna 172?" | [338, 331, 339, 726, 725] | "Based on the provided figures and diagrams for the Cessna 172, the VHF antenna is removed as follows:<br>1. **Disconnect the coaxial connector** from the base of the VHF antenna, as shown in Figures 203 (Sheet 1 and 2). This step is essential to electrically isolate the antenna from the avionics system.<br>2. **Remove the mounting screws and washers** securing the VHF antenna to the fuselage skin. Carefully remove all screws as depicted in the exploded diagrams.<br>3. **Lift off the VHF antenna**, along with the gasket if it is not adhered to the aircraft skin.<br>4. If present, **remove the gasket** from the fuselage skin.<br>5. **Inspect for any remaining hardware** (e.g., washers or vibration dampeners) and carefully remove them." |
| | MMR Ground Truth | MRAG Ground Truth |
| | [337] | 1) Set MASTER switch to OFF.<br>2) Remove the four screws and washers attaching the VHF antenna to the upper fuselage surface.<br>3) Pull antenna away from fuselage to disconnect the coax connector (PC1001 for VHF1, PC1002 for VHF2)." |

### B. MRAG Pipeline over Questioning to LLM by Uploading the Maintenance Manual

A question that may arise from this research is why this MMR and MRAG were necessary when an LLM could simply read the maintenance manual and generate a dataset, as the synthetic test dataset was created based on prompt instructions. While this is a valid point, uploading the entire document to an LLM and obtaining introduced uncertainty because the model's internal processing and reasoning were not transparent, which may produce unreliable responses. In fact, the researchers observed several typographical errors in the synthetic test dataset during the quality review process.

In contrast, the MRAG pipeline explicitly embedded the entire maintenance manual and stored it in the vector database. The MMR then retrieved manual pages based on a query, and the MRAG pipeline produced a response from those retrieved pages, resulting in outputs that were more transparent. The level of transparency is essential in safety critical domains such as aviation [41,42].

### C. Potential Benefits for Technicians

While this study did not include quantitative comparisons of actual time technicians spend searching for maintenance manual information using PDFs or physical documents, it relied solely on a synthetic test dataset for evaluation. The proposed MRAG pipeline for C172-MM demonstrated promising performance. Specifically, the MMR achieved a 92.87% recall@5, and the MRAG pipeline achieved an average BERT-based F-1 semantic similarity of 87.20%, with an average total response time of 16.88 seconds to retrieve relevant C172-MM pages and generate responses. While this study focused on C172-MM, other aircraft maintenance manuals could be developed in the future. In fact, these findings indicated that the proposed MRAG pipeline for C172-MM has the potential to implicitly reduce technicians' manual search burden, which accounts for 20 to 40% of their maintenance work [7]. It may also help technicians perform tasks that are more closely aligned with the authorized methods described in maintenance manuals.

Additionally, both novice and experienced technicians can benefit from the proposed MRAG for the C172-MM. Jo [27] reported that the RAG model reduced manual searching time for both groups, and similar advantages are expected from the MRAG pipeline. Specifically, novice technicians often lacked familiarity with detailed maintenance procedures and therefore needed to read and repeatedly refer to the maintenance manual to perform maintenance tasks [43]. For them, this multimodal retrieval approach provided relevant information more quickly, reducing the time required to locate the correct procedure and helping them follow the steps more accurately. Experienced technicians, on the other hand, may not read the maintenance manual frequently because of their familiarity and experience with most routine procedures [9]. However, when they encountered infrequently performed maintenance tasks, they needed to reference the maintenance manual. In these cases, the MMR helped them quickly access the necessary information without having to search through multiple sections of the maintenance manual.

### D. Computational Concerns in Practical Usage of MRAG Pipeline

Compared with traditional RAG model development, which required a document to undergo several pre-processing techniques [28], ColPali's page level embedding approach was more efficient during development. It significantly reduced development time and cost because it eliminated the need for extensive pre-processing and multimodal separation.

However, this efficiency came with trade-offs. As noted in ColPali model [32], page-based multimodal embeddings produced substantially larger vector representations than text-only embeddings, resulting in a significant vector database. Moreover, generating these page-based embeddings required substantially more computational resources than traditional text embedding.

If technicians wanted a textual response that directly addressed the technician's query, the retrieved pages needed to be input into the VLM, as demonstrated in the second MRAG assessment for C172-MM. However, this approach introduced additional computational cost because operating a VLM required paid resources; in this study, GPT 4.1 was used, which incurred OpenAI API token expenses. To avoid this extra cost, open source VLM could be leveraged instead.

Additionally, this research was conducted in a Google Colab CPU environment, the feasibility of deploying the proposed MRAG on portable tablet devices requires consideration of inference time. In addition, this MRAG can be deployed through the web, which requires stable and reliable network connectivity. Furthermore, aircraft maintenance manuals are periodically revised, which means that the MMR is required to be re-embedded.

## VII. Limitations and Future Work

Based on the two assessments, this research demonstrated that the MMR and the MRAG pipeline showed promising performance in retrieving C172-MM pages across all chapters and query types and in generating responses.

However, the researchers identified limitations of the ColPali model and Cessna 172-MM, along with potential improvements for future work:

## A. Limitations: ColPali Model

### *1. Page Disconnection*

A key limitation of this ColPali model is that it embeds each page as an independent image level representation. Because pages are embedded separately, the model cannot capture context that spans across page boundaries. As a result, information that continues onto subsequent pages is often retrieved in a fragmented manner.

This issue appeared when testing the MMR with the query, "*What CAUTION is issued when installing the GIA63 Integrated Avionics Unit into the mounting rack?*". The removal/installation section began on PDF page 333, followed by two figures, and then continued on PDF page 336 which contained the caution note representing the correct response. Although the caution note on page 336 was part of the same GIA63 procedure, the MMR failed to retrieve it because it treated each page as an independent unit and therefore did not recognize that these pages belonged to a continuous section.

### *2. Irrelevant Information Retrieval*

Within the top-five retrieval results, the MMR retrieved other maintenance manual pages that mentioned the GIA63 but some were irrelevant information to the query because the manual contains scattered references to the GIA63 across different sections. For example, the MMR retrieved the following pages:

The first two pages were related information to the GIA63:

1. Pdf page 333 covers about NAV/COM – maintenance practices
2. Pdf page 727 covers about GIA 63 integrated avionics installation – maintenance practices

However, the remaining three retrieved pages appeared solely due to keyword overlap, as they mentioned the GIA63 but were irrelevant to the installation procedure:

3. Pdf page 738 covers about Air data computer – maintenance practices
4. Pdf page 926 covers about Data computer – maintenance practices
5. Pdf page 704 covers about Air data computer – maintenance practices

## B. Limitations: C172-MM

*1. Lack of Textual Description Linking Content and Figures*

A failure case showed that when a query, "*Which figure in Chapter 34 shows the AHRS (GRS 77) installation and magnetometer calibration positions?*" was input into the MMR, it failed to retrieve the correct maintenance manual pages. Even though the query explicitly asked for a diagram, the MMR instead retrieved text-based procedures related to the GRS 77 AHRS, such as removal/installation steps and general maintenance practices.

Therefore, the researchers understood that the MMR assumed any AHRS GRS-77 related maintenance content was relevant, even when the query's intent was explicitly to retrieve a figure. The MMR failed to retrieve the correct figures for two reasons: first, the relevant figures were difficult to retrieve because the associated pages lacked explicit textual cues. For example, the page with the correct figure did not include the term "*AHRS (GRS-77)*" in its title; instead, the title was "*Forward Avionics Equipment Installation*."; and the procedural pages referenced the figure only indirectly, using phrases such as "*Refer to Figure 201*," without providing the actual figure title. Second, the figure did not contain any direct textual match to the query terms, such as "*Magnetometer*" or "*GRS 77*."

**C. Future Study: Potential Improvements for the ColPali model and C172-MM**

The two limitations mentioned earlier are required to be addressed to improve the performance of MMR for C172-MM. To address the first limitation, the ColPali model could be developed to better handle surrounding page retrieval when page continuation patterns are present. The second limitation relates to the structure of the C172-MM itself; restructuring the manual would help align it more effectively with ColPali's multimodal embedding approach.

In the C172-MM, page and figure references follow the ATA chapter format, resulting in repeated page and figure numbers across chapters. For example, "*figure 201*" appears in every chapter without a unique identifier, and the manual refers to figures only by number rather than by their figure titles. As a result, the MMR cannot determine which specific figure a reference corresponds to, creating ambiguity and reducing retrieval accuracy. Additionally, while the maintenance manual provides all necessary task steps, it does not specify the tools required to complete them, so technicians are required to draw on their own experience to fill these gaps. Since MRAG can only retrieve information contained in the maintenance manual, incorporating tool suggestions alongside the retrieval response could further increase the usability of the MRAG pipeline.

## VIII. Conclusions

Aircraft technicians are required to consult aircraft maintenance manuals to correctly perform maintenance. However, the usability of the maintenance manuals, particularly the difficulty of locating the correct information, often becomes a barrier. Technicians frequently need accurate information immediately, and when retrieving that information is time consuming or when the effort outweighs the perceived benefit, they may be less motivated to search thoroughly. Existing RAG models tailored to aircraft maintenance manuals are limited to retrieving textual information and cannot handle multimodal manuals, even though technicians rely heavily on visual information.

To address this gap, the researchers focused on C172-MM and developed the MMR that can retrieve maintenance manual pages and integrate the MMR into the MRAG pipeline, which can respond to a query. The MMR performance was first assessed on a synthetic test dataset of 400 queries, which produced promising results. Then, the MRAG pipeline was assessed, and the MMR's top-*k* retrieved pages were input into GPT4.1 VLM, which generated a response to the query. The researchers identified that the MRAG responses were highly semantically similar to the ground truth. In addition, three aspects of MRAG's practical feasibility were evaluated: 1. page retrieval and response generation inference times, 2. operational cost, 3. The interpretability of the MMR.

Based on these findings, the researchers conclude that the proposed MRAG pipeline has the potential to significantly reduce the time technicians spend searching through maintenance manuals by efficiently retrieving multimodal information. This improvement could streamline maintenance workflows and support closer adherence to manufacturers' specified maintenance practices by enabling technicians to access the correct procedures quickly and accurately.

## Appendix A: Zoomed-in Version of Fig. 4 Heatmap Visualizations

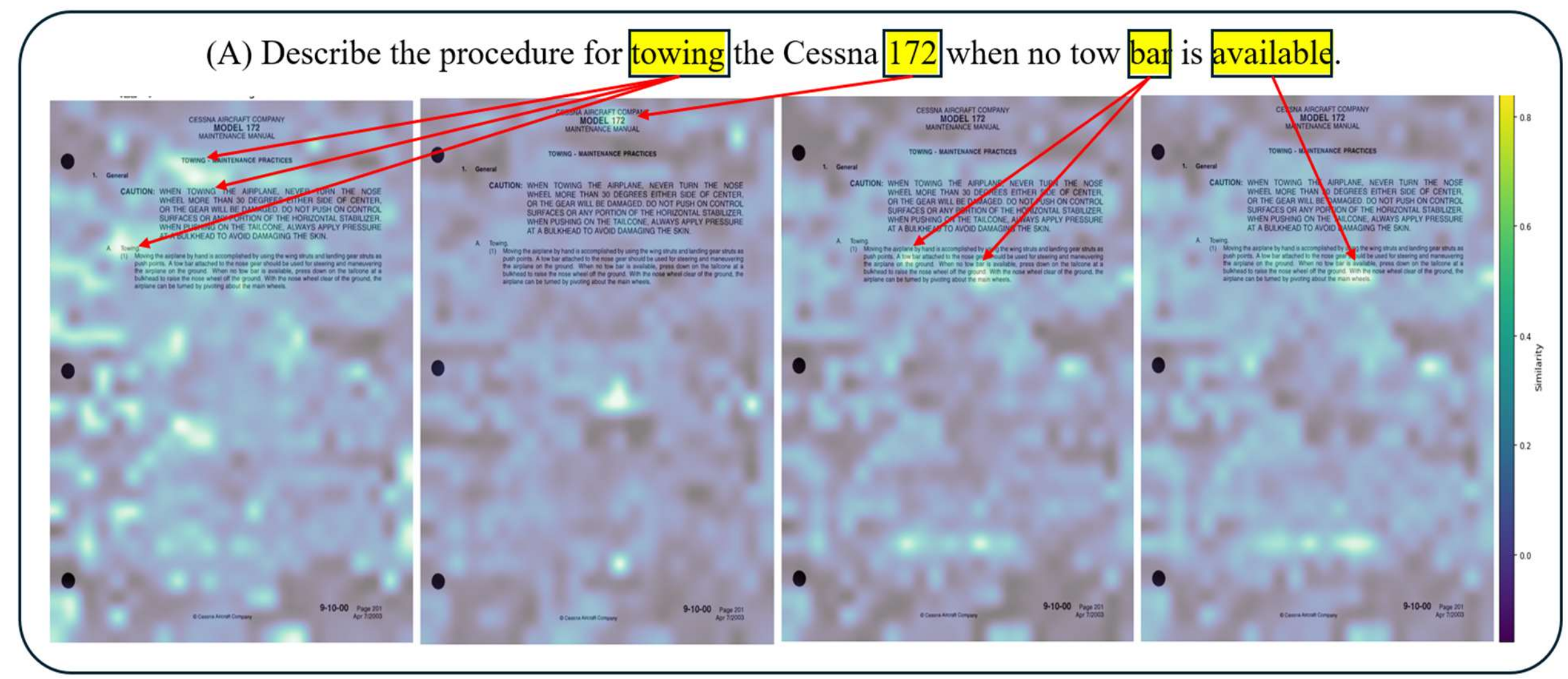

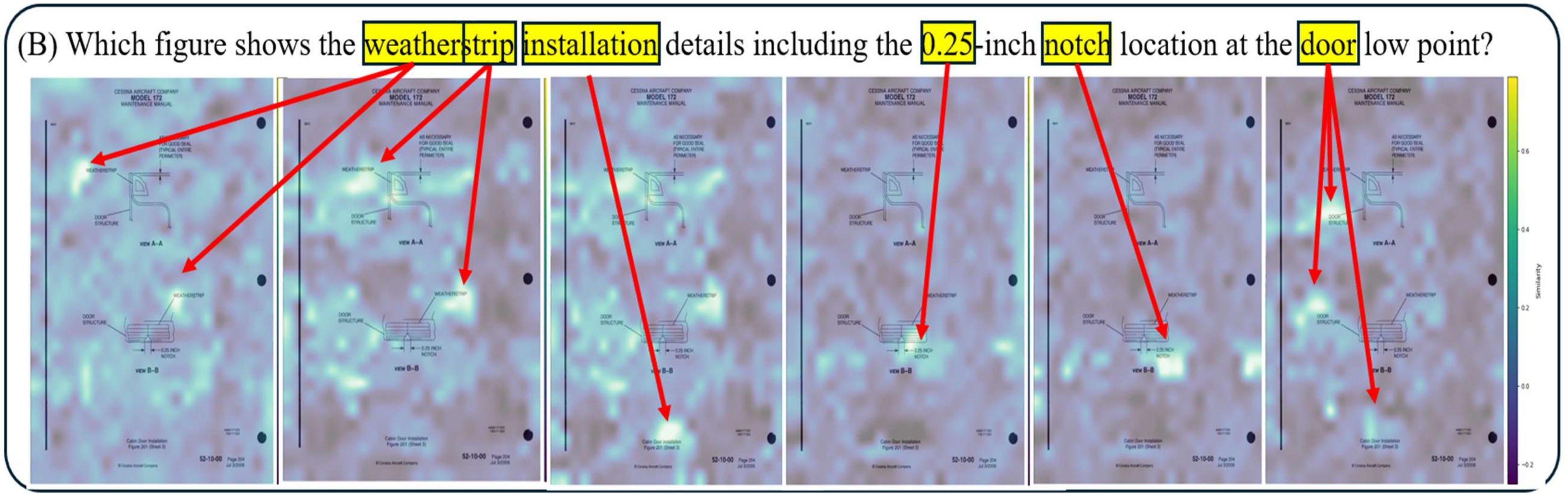

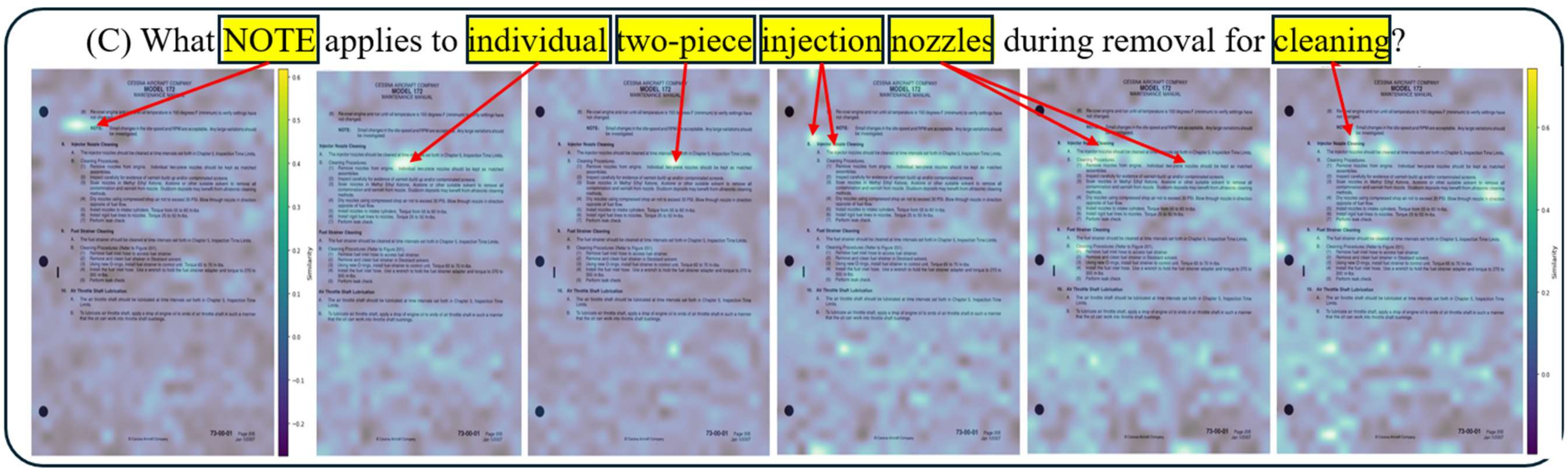
(C) What NOTE applies to individual two-piece injection nozzles during removal for cleaning?

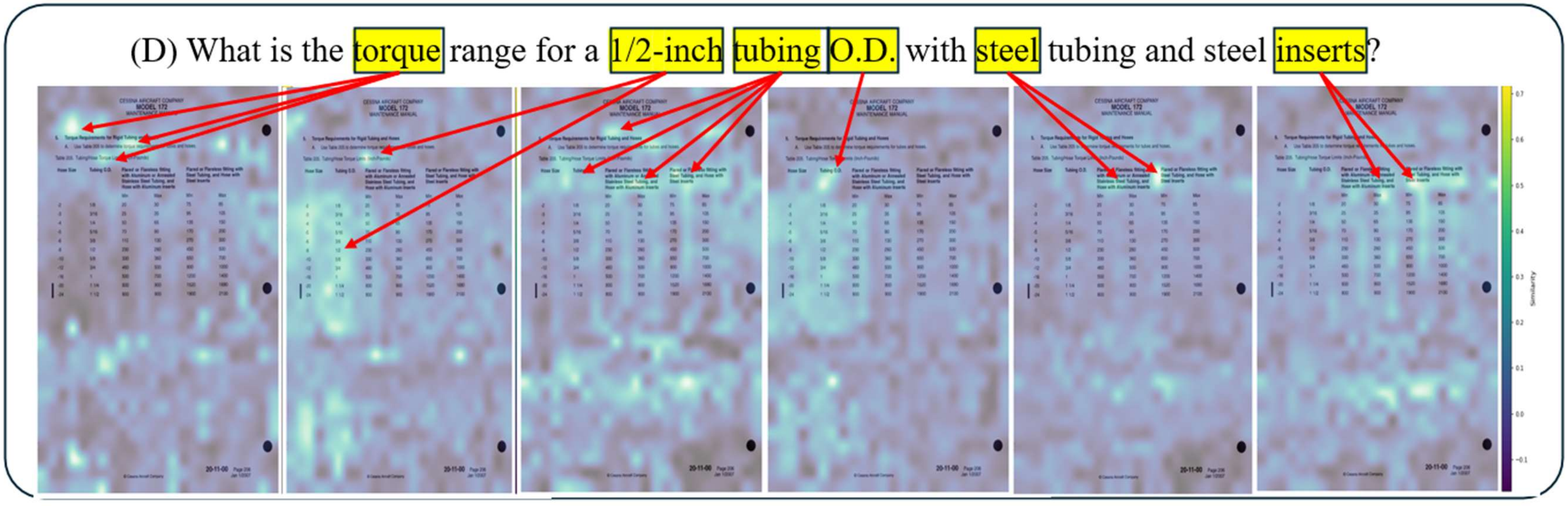
(D) What is the torque range for a 1/2-inch tubing O.D. with steel tubing and steel inserts?

## Data Availability

The dataset is not publicly available but is available from the corresponding author upon request.

## Declaration of Competing Interest

The authors declare that they have no known competing financial interests or personal relationships that could have appeared to influence the work reported in this paper.

## Declaration of Generative AI in Scientific Writing

During the preparation of this work the author(s) used Microsoft Copilot and Grammarly to refine grammar and improve the flow of the text. After using this tool/service, the author(s) reviewed and edited the content as needed and take(s) full responsibility for the content of the published article.

## Acknowledgments

This research was supported by the National Science Foundation: Future of Work at the Human-Technology Frontier, Award Abstract # 2326187